\documentclass[final] {aipproc}
\usepackage{sidecap,graphicx,boxedminipage,epsfig,wrapfig,floatflt,shadow}

\layoutstyle{8x11double}

\pdfoutput=1
\usepackage{graphicx}

\begin{document}

\title{Effect of Misconception on Transfer in Problem Solving}
\author{Chandralekha Singh}{
  address={Department of Physics and Astronomy, University of Pittsburgh, Pittsburgh, PA, 15260}}

\begin{abstract}
We examine the effect of misconceptions about friction on students' ability to solve problems and 
transfer from one context to another.
We analyze written responses to paired isomorphic problems given to
introductory physics students and discussions with a subset of students.
Misconceptions associated with friction in problems were sometimes so robust that pairing
them with isomorphic problems not involving friction did not help students fully discern their
underlying similarities.
\end{abstract}

\maketitle

\section{Introduction}

Here, we explore the use of isomorphic problem pairs (IPPs) to assess introductory physics students' ability to solve problems
and successfully transfer from one context to another in mechanics.
We call the paired problems isomorphic because they require the same physics principle to solve them.
We examine the effect of misconceptions about friction as a potential barrier for problem solving and
analyze the performance of students on the IPPs from the perspective of ``transfer".~\cite{gick,mestre,lobato,sanjay}
Transfer in physics is particularly challenging because there are only a few principles and concepts that are condensed
into a compact mathematical form. Learning requires unpacking them and understanding their applicability in a variety of 
contexts that share deep features, e.g., the same law of physics may apply in different contexts.
Cognitive theory suggests that transfer can be difficult
especially if the ``source" (from which transfer is intended) and the ``target" (to which transfer is intended)
do not share surface features. This difficulty arises because
knowledge is encoded in memory with the context in which it was acquired and solving the source problem
does not automatically manifest its ``deep" similarity with the target problem.~\cite{gick}
Ability to transfer improves with expertise because an expert's knowledge is hierarchically organized and represented
at a more abstract level in memory, which facilitates 
categorization and recognition based upon deep features.~\cite{mestre,lobato,sanjay}

In this investigation, students in college calculus-based introductory physics courses were given IPPs 
in the multiple-choice format with one problem in each pair involving friction 
to examine the effect of misconceptions about friction on problem solving and transfer. One common misconception
about the static frictional force is that it is always at its maximum value because students have difficulty with the 
mathematical inequality that relates the magnitude of the static frictional force with the normal force.~\cite{disessa}
Students also have difficulty determining the direction of frictional force.
If students are given IPPs in which one problem has distracting features such as common
misconceptions related to friction, students may have difficulty transferring from the problem that did not have
the distracting features to the one with distracting features.
We analyze why it is difficult for students to fully discern the deep similarity of the problems in such IPPs and apply the
strategies they successfully used in one problem to its pair.

In selecting and developing questions for the IPPs, we used our prior experience; one of the questions of each IPP was
a question that had been found difficult by introductory physics students previously.
Some students were given both problems of an IPP while others were given only one of the two problems.
Students who were given both problems of an IPP were \underline{not} told explicitly that the problems given were isomorphic.
In some cases, depending upon the consent of the course instructor (due to time constraint for a class),
students were asked to explain their reasoning in each case to obtain full credit. The problems
contributed to students' grades in all courses.
In some of the courses, we discussed the responses individually with a subset of students.

\vspace*{-.1in}
\section{Discussion}
\vspace*{-.1in}

We find that students have misconceptions about friction~\cite{disessa} that prevented
those who were given both problems of the IPPs from taking advantage of the problem not involving
friction (which turned out to be easier for them) to solve the paired problem with friction.
Questions (1)-(4) are related to an IPP about a car in equilibrium on an incline 
which are equilibrium applications of Newton's second law. 
Students have to realize that the car is at rest on the incline in each case, so the net force on
the car is zero. Also, since the weight of the car and the normal force exerted on the car by the inclined surface are
the same in both problems, the only other force acting on the car (which is the tension force in one problem and the static frictional force
in the other problem) must be the same. The correct answer (the magnitudes of the tension and frictional forces)
in the problems is $7,500$ N, equal to the component of the weight of the car acting down the incline. The correct
responses in the questions below have been italicized:\\

\noindent
Note: These trigonometric results might be useful in the next four questions:
$\sin 30^0=0.5,\,\,\,\,\,\,\,$ $\cos 30^0=0.866$.

\vspace{0.02in}

\noindent
{\bf {$\bullet$} \underline{Setup for the next two questions}}\\

\noindent
{\bf A car which weighs $15,000$ N is at rest on a frictionless $30^0$ incline,
as shown below. The car is held in place by a light strong cable parallel to the incline.}

\begin{center}
\includegraphics[width=1.31in]{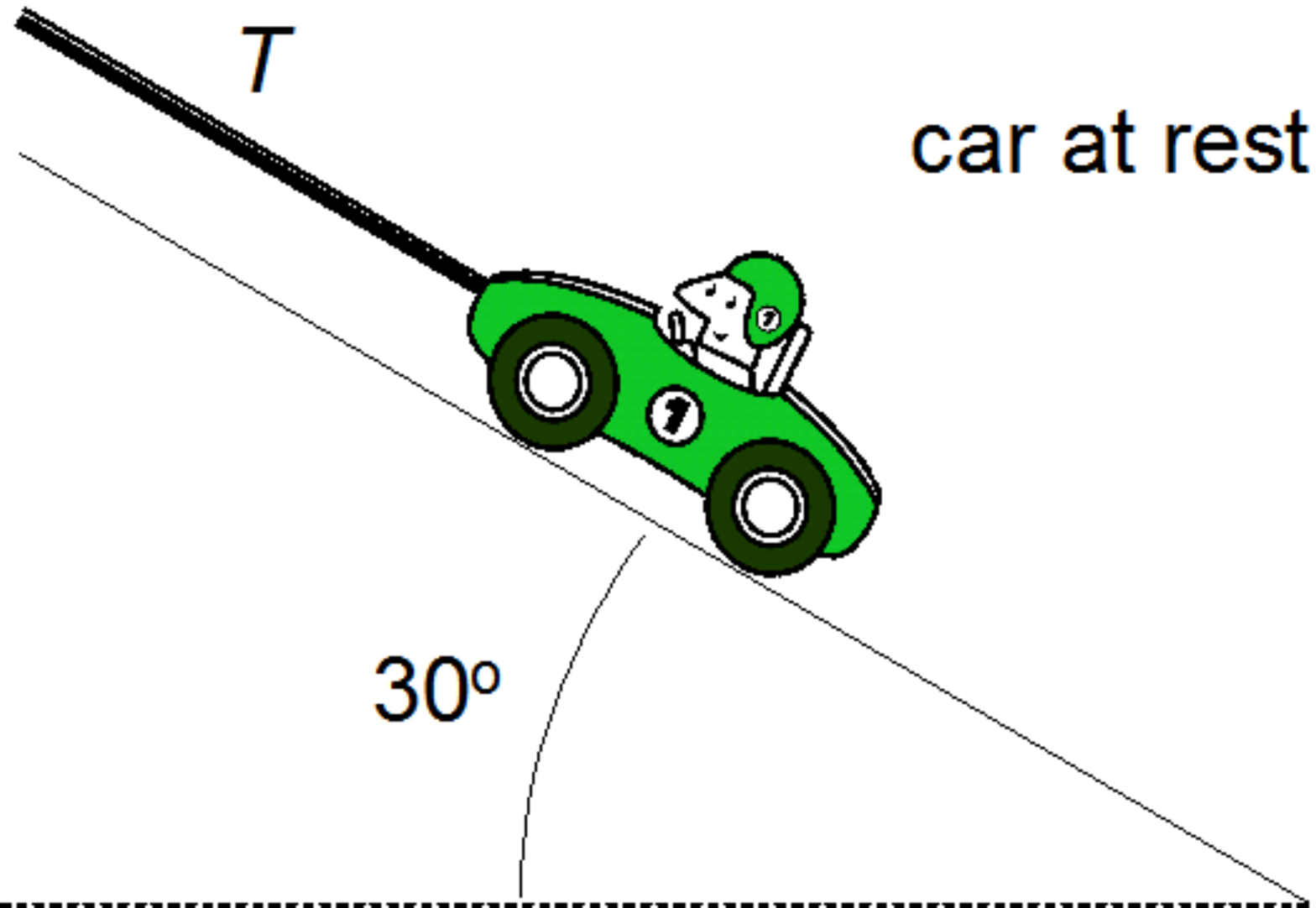}

\end{center}

\noindent
(1) Which one of the following is the correct free body diagram of the car (with correct directions)?
$N$, $mg$, and $T$ are the magnitudes of the normal force,
the weight of the car and the tension force, respectively.\\
\begin{center}
\vspace*{-.23in}
\includegraphics[width=2.31in]{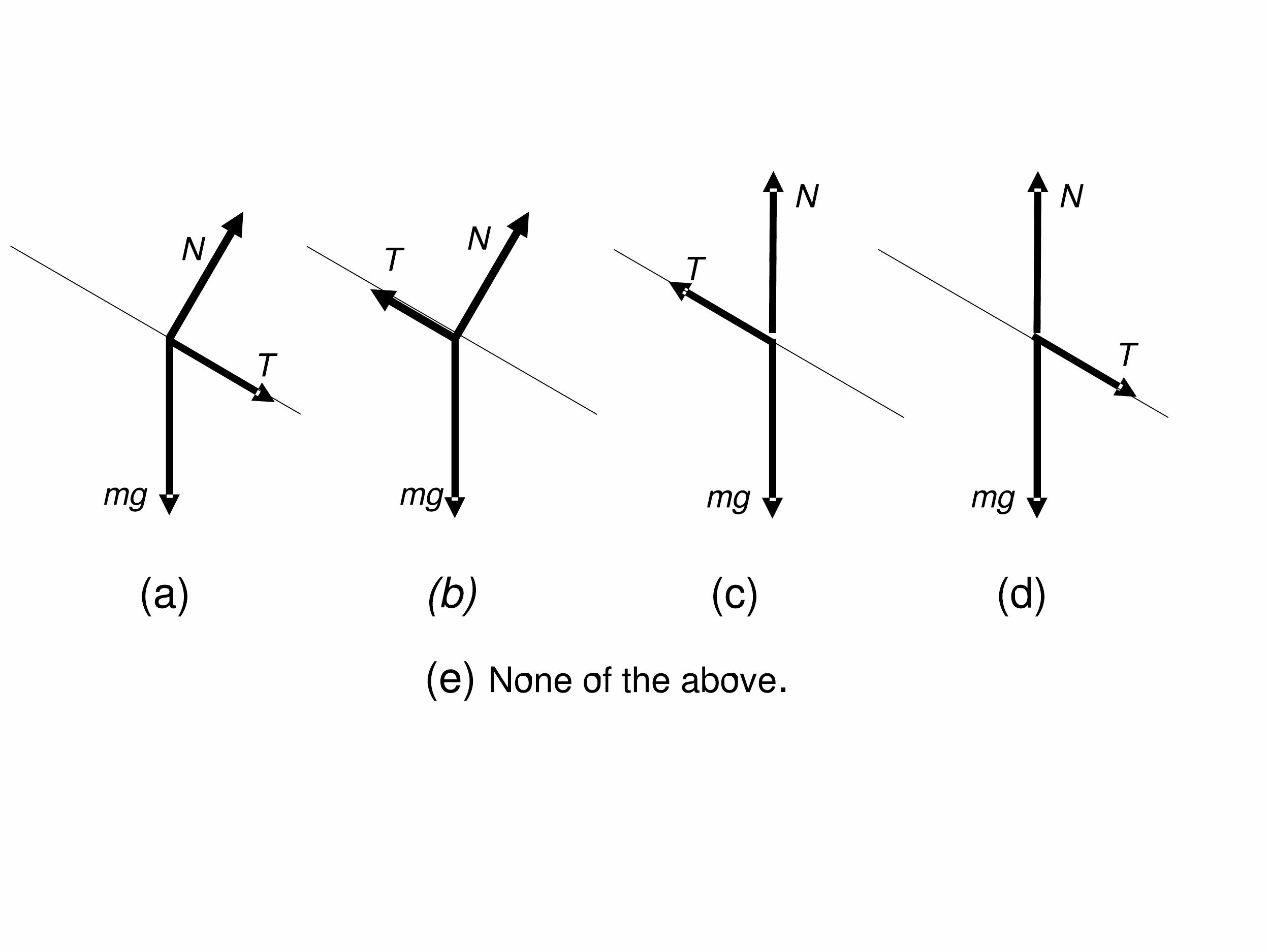}
\vspace*{-.18in}
\end{center}

\noindent
(2) Find the magnitude of tension force $\vec T$ in the cable.\\
{\bf {\it (a) $7,500$ N}}\\
(b) $10,400$ N\\
(c) $11,700$ N\\
(d) $13,000$ N\\
(e) $15,000$ N

\vspace{0.03in}

\noindent
{\bf {$\bullet$} \underline{Setup for the next two questions}}\\

\noindent
{\bf A car which weighs $15,000$ N is at rest on a $30^0$ incline, as shown below.
The coefficient of static friction between the car's tires and the road is $0.90$, and
the coefficient of kinetic friction is $0.80$.}

\begin{center}
\includegraphics[width=1.222in]{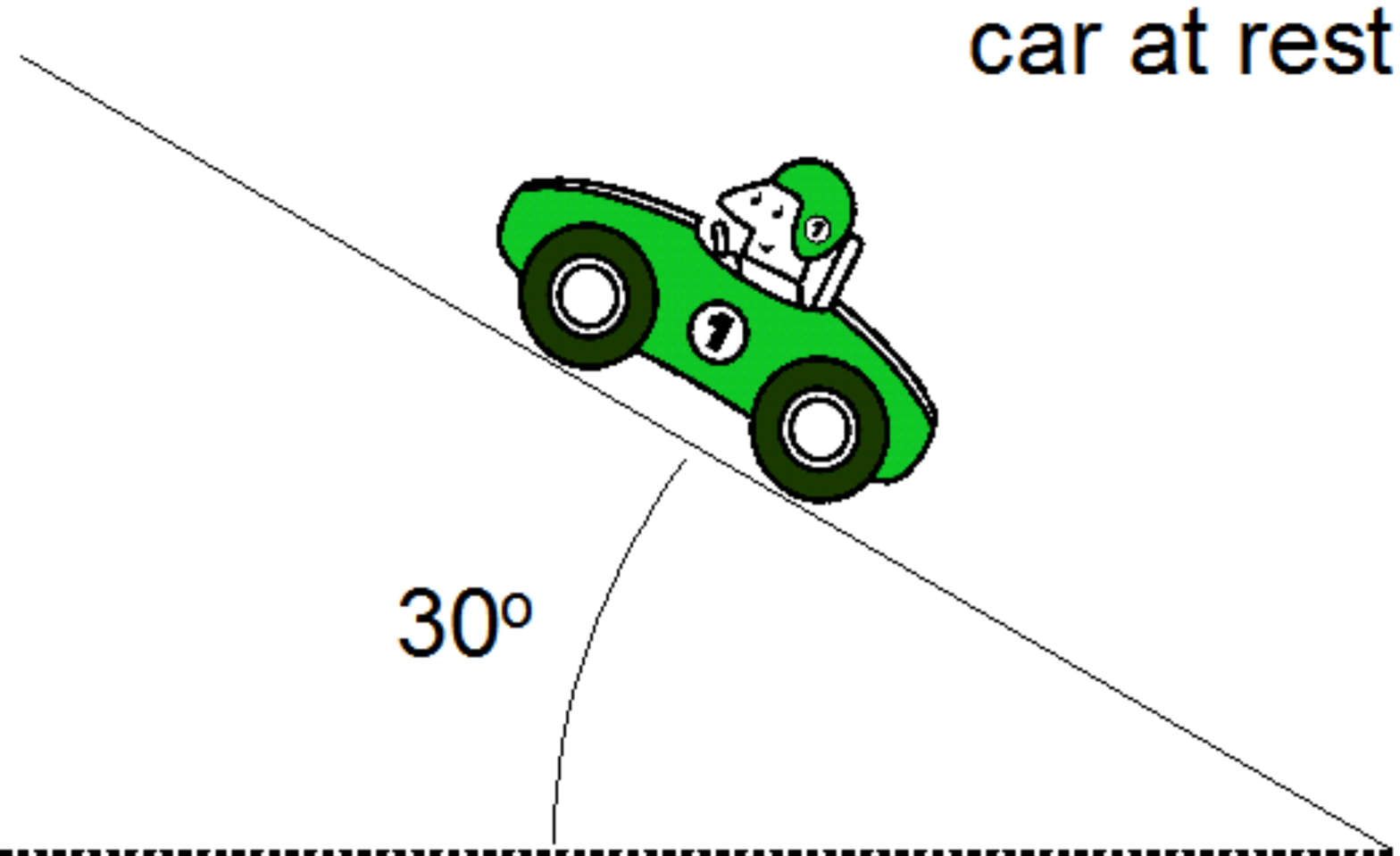}

\end{center}

\noindent
(3) Which one of the following is the correct free body diagram of the car (with correct directions)?
$N$, $mg$, and $f_s$ are the magnitudes of the normal force,
the weight of the car and the static frictional force, respectively.\\
\begin{center}
\vspace*{-.22in}
\includegraphics[width=2.19in]{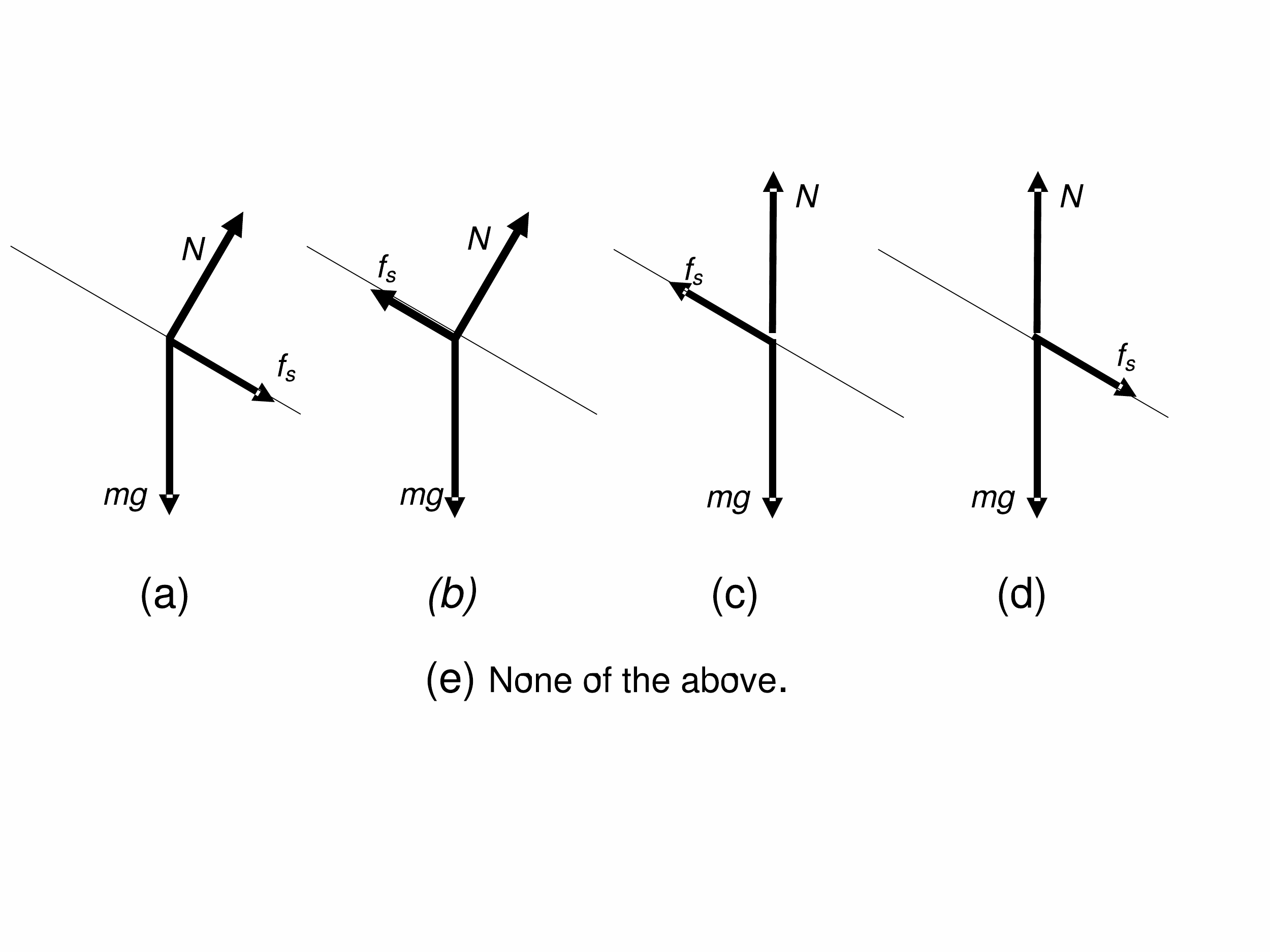}

\vspace*{-.18in}
\end{center}

\noindent
(4) Find the magnitude of the frictional force on the car.\\
{\bf {\it (a) $7,500$ N}}\\
(b) $10,400$ N\\
(c) $11,700$ N\\
(d) $13,000$ N\\
(e) $15,000$ N

One common misconception about the static frictional force is that it must be at its maximum value $f_s^{max}=\mu_s N$
where $\mu_s$ is the coefficient of static friction and $N$ is the magnitude of the normal force.
We wanted to investigate if student
performance was significantly different for the cases in which students worked {\it only} on the problem
with friction vs. when that problem is preceded by the problem involving tension in
the cable. We wanted to explore if students in the latter group observed the underlying similarity of the IPPs
and realized that the static frictional force was less than its maximum value and exactly identical to the tension in the
cable in the problem pair.

First, 115 students were only asked question (4) and a different 272 students
were asked both questions (2) and (4), given in that order. Questions (1) and (3) about the free body
diagrams were not given to either of these groups. Out of the 115 students, $16\%$ provided a correct response when only
question (4) was asked. Out of the 272 students, $61\%$ provided a correct response for question (2) with the tension
force holding the car at rest but only $17\%$ provided a correct response for the case with friction.
The two most common incorrect responses in question (4) were $\mu_s N=11,700$ N  ($\sim 40\%$) and $\mu_k N=10,400$ N ($\sim 20\%$).
Thus, giving both problems in the IPP did not improve student performance on the problem with friction.
In other words, the misconception
about friction dominated and most students who were given both problems did not discern the underlying similarity of the 
two problems. 

In order to help students discern the similarity between questions (2) and (4), we later introduced two additional questions that
asked students to identify the correct free body diagrams for each of the questions. We wanted to assess whether forcing students to think
about the free body diagram in each case helps them focus on the similarity of the problems. Forty-five students were given questions
(1)-(4).
We find that $89\%$ and $69\%$ of the students identified the correct free body diagrams in questions (1) and (3), respectively.
The most common incorrect response ($\sim 25\%$) in question (3) was choice (a) because these students believed that the frictional force
should be pointing down the incline. Unfortunately, the percentage of correct responses for questions (2) and (4) were $69\%$ and $14\%$,
respectively, not statistically different from the corresponding percentages without the corresponding free body diagram questions. Thus, 
although $89\%$ and $69\%$ of the students identified the correct free body diagrams in questions (1) and (3), it did not help 
them see the similarity of questions (2) and (4) or challenge their misconception about the frictional force 
(again $\sim 40\%$ believed
that friction had a magnitude $\mu_s N$ and $\sim 30\%$ believed it was $\mu_k N$).
In individual interviews, students often noted that the problem with friction must be solved differently from the problem
involving tension because there is a special formula for the frictional force. Even when the interviewer drew students' attention
to the fact that the other forces (normal force and weight) were the same in both questions and they are both equilibrium problems, only some of the students
appeared concerned. Others used convoluted reasoning and asserted that friction has a special formula which should
be used whereas tension does not have a formula, and therefore, the free body diagram must be used.

The IPP in questions (5) and (6) below relates to the work done by a person while pushing a box over
the same distance at a constant speed parallel to different ramps with or without friction.
Although the frictional force in question (6) is irrelevant for the question asked, it was a distracting feature for a majority
of students:

\vspace{0.06in}

(5) You are working at a bookstore. Your first task is to push a
box of books $1.5$ m along a frictionless ramp at a constant speed.
You push parallel to the ramp with a steady force of
$500$ N. Find the work done on the box {\bf by you}.\\
(a) $200$ J\\
(b) $300$ J\\
{\bf {\it (c) $750$ J}}\\
(d) $1000$ J\\
(e) Impossible to calculate without knowing the angle of the ramp.

\vspace{0.01in}

\noindent
(6) You are working at a bookstore. Your second task is to push a
box of books $1.5$ m along a rough ramp at a constant speed.
You push parallel to the ramp with a steady force of
$500$ N. A frictional force of $300$ N opposes your efforts. Find the work done on the box {\bf by you}.\\
(a) $200$ J\\
(b) $300$ J\\
{\bf {\it (c) $750$ J}}\\
(d) $1000$ J\\
(e) Impossible to calculate without knowing the angle of the ramp.

\vspace{0.01in}

Out of the 131 students who answered both
questions, $65\%$ and $27\%$ provided the correct responses to questions (5) and (6), respectively.
Common incorrect reasonings for question (6) was assuming that friction must play a role in determining the work done by the
person and
the angle of the ramp was required to calculate this work even though the distance by which the box was moved along the ramp was given.
Interviews suggest that many students had difficulty distinguishing between
the work done on the box by the person and the total work done.
They asserted that the work done by the person cannot be the same in the two problems because
friction must make it more difficult for the person to perform the work.

In another IPP related to the equilibrium of a box or a table on a horizontal surface (questions (7) and (8) below), 
students had to realize that since the net force on the object is zero, the force exerted by you and Arnold on the box must be equal in
magnitude in question (7) and the force exerted by you and the frictional force on the table must be equal in magnitude in question (8):

\vspace{0.04in}

\noindent
(7) Arnold and you are both pulling on a box of mass $M$ that is at rest on a frictionless surface, as
shown below. Arnold is much stronger than you. You pull horizontally as hard as you can, with a force $\vec f$, and
Arnold keeps the mass from moving by pulling horizontally with a force $\vec F$. Which one of the following is a correct
statement about the magnitude of Arnold's force $\vec F$? $g$ is the magnitude of the acceleration due to gravity.\\
\begin{center}
\includegraphics[width=1.0in]{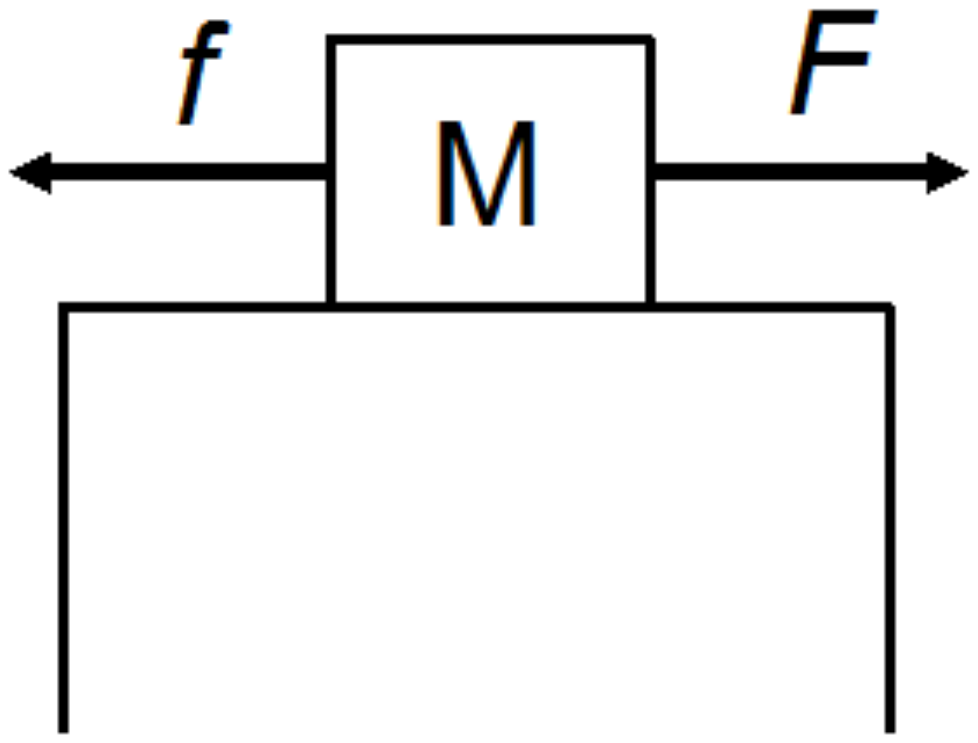}
\end{center}

\noindent
{\bf {\it (a) $F=f$ because the mass is not accelerating.}}\\
(b) $F=f+Mg$ because Arnold must balance out your force and the weight.\\
(c) $F>f$ because Arnold is stronger and pulls harder.\\
(d) $F<f$ because Arnold is stronger and need not pull as hard as you.\\
(e) $F+f=Mg$ to maintain equilibrium.\\

\noindent
(8) You are trying to slide a table across a horizontal floor. You push horizontally on the table
with a force of $400$ N. The table does not move.
What is the magnitude of the frictional force the rug exerts on the table?
The coefficient of static friction between the table and the rug is $0.60$,
and the coefficient of kinetic friction is $0.50$. The table's weight is $1000$ N.  \\
(a) $0$ N\\
{\bf {\it (b) $400$ N}}\\
(c) $500$ N\\
(d) $600$ N\\
(e) $1000$ N\\

We asked 115 students only question (8) while a different 272 students were asked both questions (7) and (8), given in that order.
Out of the 115 students, $22\%$ provided a correct response when only question (8) was asked. Out of the 272 students, $69\%$
provided a correct response for question (7) and only $24\%$ for question (8).
The most common incorrect response in question (8) was $\mu_s N=600$ N ($\sim 40\%$) with or without question (7).
Similar to the car on the inclined plane problem, the misconceptions about friction were so strong that students who were given
both problems did not fully discern their similarity and take advantage of their response to question (7) to analyze
the horizontal forces in question (8).
To understand the extent to which students have difficulty with the magnitude and direction of the static frictional force, 131 students
were also given question (9) below:

\vspace{0.1in}

\noindent
(9) A packing crate is at rest on a horizontal surface. It is acted on by three horizontal forces:
$600$ N to the left, $200$ N to the right and friction. The weight of the crate is $400$ N. If the $600$ N
force is removed, the resultant force acting on the crate is\\
{\bf {\it (a) zero}}\\
(b) $200$ N to the right\\
(c) $200$ N to the left\\
(d) $400$ N to the left\\
(e) impossible to determine from the information given.\\

The correct response for question (9) is (a), i.e., there is no resultant force acting
on the crate because the $200$ N force to the right will be balanced by a $200$ N static frictional force in the opposite direction.
Only $15\%$ of the students provided the correct response. In question (9), the coefficient of friction was not provided, and
similar to the common misconception in question (4), almost $50\%$ of the students believed that it is impossible to determine the resultant force on the crate without this information.

\vspace*{-.2in}
\section{Summary and Conclusions}
\vspace*{-.1in}

We explored the effect of misconceptions about friction on problem solving and transfer, 
and find that they interfere with the ability to reason correctly. 
The responses to the IPPs involving friction shows that students had difficulty in seeing the deep connection between the IPPs
and in transferring their reasoning from the problem not involving friction to the problem with friction.
The fact that students did not take advantage of the problem in an IPP not involving friction (which turned out
to be easier for them) to answer the 
corresponding problem involving friction suggests that the misconceptions about friction were quite robust.
For example, many students believed that the static friction is always at the maximum value or the kinetic friction
is responsible for keeping the car at rest on an incline or the presence of friction must affect the work done by you
(even if you apply the same force over the same distance).
Even asking students to draw the free body diagrams explicitly in some cases did not help most students.

In such cases where misconceptions about friction prevented transfer, students may benefit from paired problems only after they are
provided opportunity to repair their knowledge structure so that there is less room for misconception.
Instructional strategies embedded within a coherent curriculum that force students to realize, e.g., that the static frictional
force does not have to be at its maximum value or that the work done by a person who is applying a fixed force over a fixed distance
cannot depend on friction may be helpful. Eliciting student difficulties by asking them to predict what should happen
in concrete situations, helping them realize the discrepancy between their predictions and what actually happens, and then
providing guidance and support to enhance their expertise is one such strategy.

\vspace*{-.12in}
\begin{theacknowledgments}
We are grateful to the NSF for award DUE-0442087.
\end{theacknowledgments}
\vspace*{-.10in}

\bibliographystyle{aipproc}

\end{document}